# Intrinsic Resistive Switching in Microtubule-Templated Gold Nanowires for Reconfigurable Nanoelectronics

*Borja Rodriguez-Barea, Brenda Palestina Romero, Charlotte Kielar, René Hübner, Stefan Diez*, Artur Erbe**

Borja Rodriguez-Barea

Institute of Ion Beam Physics and Materials Research, Helmholtz-Zentrum Dresden-Rossendorf, 01328 Dresden, Germany

Brenda Palestina Romero

B CUBE - Center for Molecular Bioengineering, TUD Dresden University of Technology, Germany, 01307 Dresden, Germany

ORCID: 0000-0003-1504-798X

Charlotte Kielar

Institute of Ion Beam Physics and Materials Research, Helmholtz-Zentrum Dresden-Rossendorf, 01328 Dresden, Germany

ORCID: 0000-0002-3002-6098

René Hübner

Institute of Ion Beam Physics and Materials Research, Helmholtz-Zentrum Dresden-Rossendorf, 01328 Dresden, Germany

ORCID: 0000-0002-5200-6928

Stefan Diez

B CUBE - Center for Molecular Bioengineering, TUD Dresden University of Technology, Germany, 01307 Dresden, Germany

Cluster of Excellence Physics of Life, TUD Dresden University of Technology, 01307 Dresden, Germany

Max Planck Institute of Molecular Cell Biology and Genetics, 01307 Dresden, Germany

ORCID: 0000-0002-0750-8515

E-mail: stefan.diez@tu-dresden.de

Artur Erbe

Institute of Ion Beam Physics and Materials Research, Helmholtz-Zentrum Dresden-Rossendorf, 01328 Dresden, Germany

TUD Dresden University of Technology, 01062 Dresden, Germany

ORCID: 0000-0001-6368-8728

E-mail: a.erbe@hzdr.de

## Funding

B.R.B., B.P.R., S.D. and A.E. acknowledge financial support funded by the Deutsche Forschungsgemeinschaft (DFG, German Research Foundation) - GRK 2767 - Project number 451785257.

The authors acknowledge the funding of TEM Talos by the German Federal Ministry of Education and Research (BMBF, grant No. 03SF0451) in the framework of HEMCP.

## Keywords

## Abstract

The scaling limitations of conventional transistors demand alternative device concepts capable of dynamic reconfigurability at the atomic scale. Resistive switching (RS), a key mechanism for neuromorphic computing and non-volatile memory, has been widely demonstrated in oxides, semiconductors, and nanocomposites, but not in pure one-dimensional metallic systems. Here, we report the first electrical characterization of gold nanowires (AuNWs) synthesized within the lumen of functionalized microtubules. Structural analyses confirm continuous metallic AuNWs with local compositional inhomogeneities. Electrical measurements reveal three distinct conduction behaviors and abrupt resistance transitions under applied bias, consistent with electromigration driven by structural reorganization. Millisecond voltage pulsing enables active and reproducible modulation of resistance states without loss of metallic conduction, establishing a new RS mechanism intrinsic to pure metallic nanowires. Owing to their high aspect ratio, lateral geometry, and CMOS-compatible processing, microtubule-templated AuNWs provide a versatile platform for reconfigurable interconnects.

## Contributions

Borja Rodriguez-Barea and Brenda Palestina Romero contributed equally to this work.

## 1. Introduction

In the coming years, next-generation technologies will demand electronic devices with unprecedented speed, efficiency, and adaptability. To counteract the decline of Moore's law,[1] precise atomic-scale engineering of materials and device structures has become a critical challenge. Despite steady improvements in processor frequency over the last decade,[2] conventional transistors are increasingly constrained by quantum effects, challenging both performance and reliability. This underscores the need for devices capable of dynamic reconfigurability through atomic and nanoscale control of their structure and properties.

To address the shortcomings of traditional static computing paradigms,[3,4] alternative computing architectures are currently being explored. Neuromorphic computing[5] addresses the data throughput disparity between the CPU (processing unit) and memory (storage), referred to as the Von-Neumann bottleneck, which is expected to be magnified with the large data consumption of artificial intelligence tasks. A promising route toward neuromorphic systems relies on resistive switching (RS), a mechanism in which a material's resistance changes dynamically in response to an applied electrical current in a non-volatile and reversible manner.

RS has been reported in a wide range of materials, including oxides, nitrides, semiconductors, and organic materials,[6,7] as well as in networks of nanoparticles (NPs) and nanowires (NWs), often incorporating insulating components within a metallic polymeric matrix or passivating them via a shell of ligands or oxide layers.[8] In metallic NWs specifically, RS has been observed in isolated metal-oxide NWs,[9,10] sandwiched oxide-metal NWs between metallic contacts[11–14], and NWs with a metallic core inside an oxide shell. [15] Thanks to their low working voltage and cycling endurance, some of these materials have been proposed for Resistive Switching Memory (RSM), relying on the rupture of conducting filaments formed in the material that change its resistance.[16] Particularly, RS behavior has not yet been demonstrated in pure one-dimensional metallic nanowire systems.

Attempts to model RS phenomena in low-dimensional materials have primarily focused on network-based systems.[7] For other nanostructured materials, such as nanogranular films, two challenges are central: identifying the microscopic mechanisms underlying discrete resistance jumps and reliably estimating the overall film resistance in both high- and low-conductance regimes.[17] Conventional models derived from homogeneous crystalline materials fail to capture the inhomogeneity and disorder inherent to nanogranular systems, limiting their predictive power. To better describe such systems, electronic transport must be treated in conjunction with microstructural evolution - a perspective that could also shed light on RS mechanisms in granular nanowires. Granularity in metallic nanowires has, for example, been observed in bio-templated metallic nanowires grown from small NPs.[18–26]

In this context, we have recently focused on microtubule-lumen-templated platforms that enable the synthesis of quasi-1D gold nanowires (AuNWs). Microtubules are cytoskeletal filaments composed of α- and β-tubulin, which assemble into hollow cylinders with outer and inner diameters of 25 nm and 15 nm, respectively. They can serve as templates for metallic nanowires both on their outer surface[22,27] and within their inner lumen.[20] Recent reports suggest that functionalizing the lumen enhances the control, uniformity, structural integrity, and conductivity of the resulting hybrid nanostructures.[20,21] In particular, high-aspect-ratio AuNWs can be grown by first attaching 1.4 nm gold NPs (AuNPs) inside the lumen, which then grow into continuous wires upon addition of a gold precursor and reducing agent.[20]

Here, we report the first electrical characterization at both room and low temperatures of AuNWs synthesized within microtubule lumens. Measurements on 50 conducting AuNWs revealed three distinct types of electrical behavior, associated with structural defects but showing no clear dependence of resistance on wire length. Under applied bias, abrupt resistance transitions were observed, consistent with a RS behavior driven by structural evolution in pure metallic nanowires. These findings establish microtubule-templated AuNWs as promising candidates for RS-based memory and logic devices, owing to their high aspect ratio, structural integrity and precisely controlled nanoscale dimensions.

# 2. Results and Discussion

To fabricate microtubule-templated AuNWs, the lumen of GMP-CPP-stabilized microtubules were labelled with functionalized AuNPs.[20] Upon addition of a gold precursor ($HAuCl_4$) and a reducing agent ($NH_2OH$), the AuNWs were templated by the microtubule lumen (**Figure 1A**, upper panel). Transmission electron microscope (TEM) images of AuNWs revealed a mean length of 176 nm and a mean diameter of 17 nm (**Figure 1B** and **Figure S1**). Then, the same samples were drop-cast on plasma-activated $SiO_2$ substrates containing positional markers (density ≈ 1 *AuNW*/50 $\mu m^2$). The markers on the $SiO_2$ substrates helped to extract the position of selected AuNWs for further processing. Afterwards, $O_2$ plasma treatment was performed to remove the organic microtubule parts, leaving clean AuNWs behind for the nanofabrication of electrical contacts on them. Cross-sectional TEM images were acquired from single AuNWs (after $O_2$ plasma treatment) in order to observe atomic arrangement and composition along the AuNWs (**Figures 1C** and **1D** and **Figure S7**). These images reveal structural and compositional non-uniformities across the AuNWs, suggesting that local variations in crystallinity govern their resistive switching behavior. Additionally, no surrounding biotemplate or organic ligand was identified from the element distribution maps. Finally, the contacts were fabricated by electron beam lithography (EBL), using an optimized version of a previously described approach.[23]

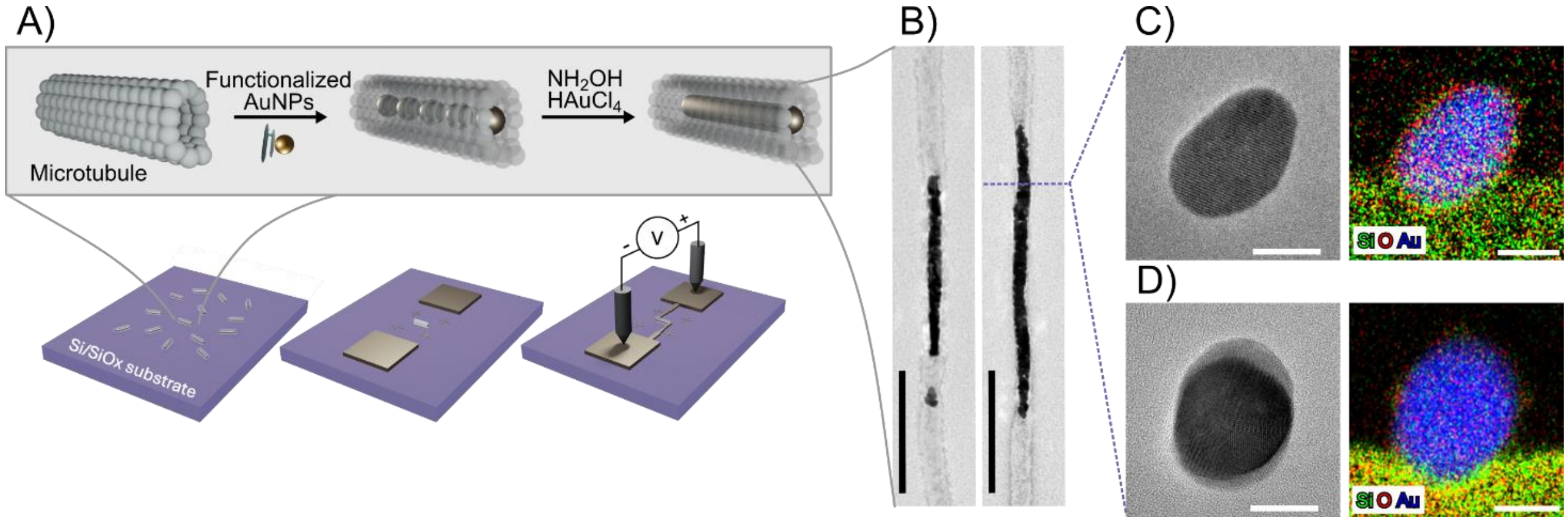

**Figure 1. A)** Schematic diagram of the synthesis and electrical characterization of microtubule-lumen-templated AuNWs. The AuNWs are deposited on a $SiO_2$ substrate. Gold references are fabricated to mark the positions of the AuNWs. Then, the organic microtubule template is removed by oxygen plasma treatment, and selected AuNWs are contacted by lithographically patterned gold electrodes for electrical measurements. **B)** TEM images of microtubule-lumen-templated AuNWs after synthesis (scale bars = 200 nm). **C)** and **D)** Cross-sectional TEM views of AuNWs without microtubules (left column) and compositional analysis showing silicon in green, oxygen in red, and gold in blue (right column, scale bars = 10 nm).

Scanning Electron Microscopy (SEM) imaging of the deposited AuNWs on the $SiO_2$ substrates before and after being contacted shows that they adhere strongly to the substrate and retain their integrity throughout the lithography process (**Figure 2**, two left columns). Although in these images the AuNWs appear to consist of continuous metal, it is only through electrical measurements that their electrical properties can be determined. Initial 2-electrode measurements conducted at room temperature and under high vacuum ($P \approx 1 \times 10^{-5}$ mbar) revealed a wide variation in resistance values, as can be observed in the *I-V* curves (**Figure 2**, right column and **Figure S2**). Most of the measured AuNWs exhibited linear characteristics (the slope of the curve determining the resistance) in a range of the source-drain voltage $V_{sd}$ from −15 mV to 15 mV. We then classified the AuNWs into three groups: (i) resistive AuNWs (with resistances from $1 \times 10^4$ Ω to $1 \times 10^9$ Ω), (ii) conductive AuNWs (with resistances below $1 \times 10^4$ Ω), and (iii) nonlinear AuNWs where about 30% of the AuNWs (15 out of 50) exhibited nonlinear *I-V* curves upon applying voltages in the range from −1 V to 1 V (with resistances from $1 \times 10^9$ Ω to $1 \times 10^{12}$ Ω, estimated from the current at the largest voltage value).

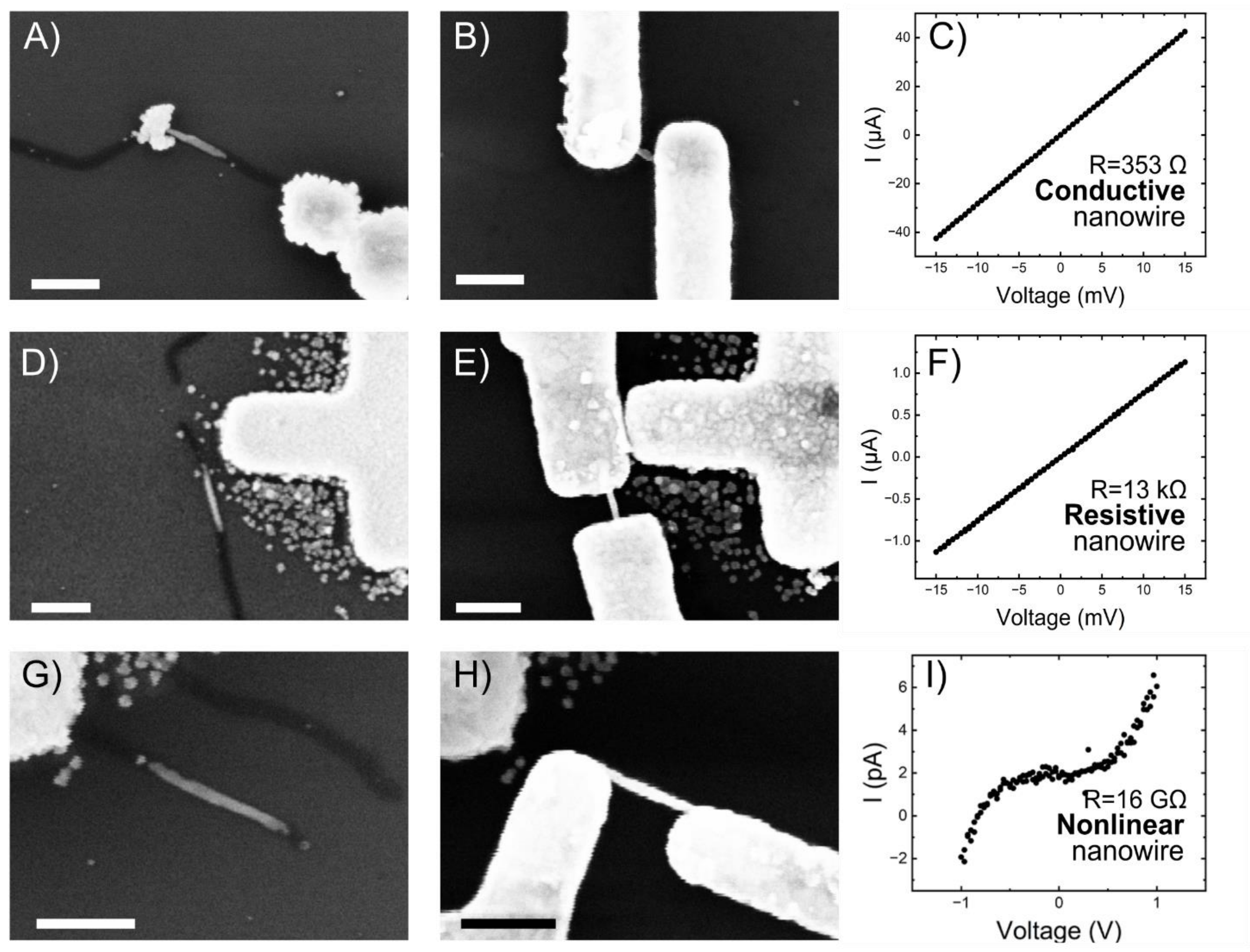

**Figure 2.** SEM images and *I-V* curves of microtubule-templated AuNWs. SEM images of deposited AuNWs. **A, D, G)** before, and **B, E, H)** after connecting them to nanofabricated electrodes. **C, F, I)** Two-electrode *I-V* curves for a conductive AuNW (C), a resistive AuNW (F), and a nonlinear AuNW (I). Measurements were performed at room temperature and under high vacuum. Scale bars = 500 nm.

The extracted resistance values are not correlated with the lengths of the AuNWs for either linear or nonlinear AuNWs (**Figure 3**). For linear AuNWs, similar characteristics have been observed in DNA-origami-templated AuNWs.[18,23,24] The conductance variability could be attributed to variations in electrode-AuNW contact geometry interfaces and intrinsic inhomogeneities among AuNWs such as discontinuities in the lattice, the presence of impurities (e.g. adsorbed molecules), grain boundaries or differences in their geometrical features. The latter are confirmed by the cross-sectional TEM-based analysis, see **Figures 1C** and **1D**).

In case of nonlinear AuNWs, the shape of the *I-V* curves may be explained by the presence of gaps or impurities inside the AuNWs. Gaps in self-assembled AuNWs lead to transport through insulating materials or tunnelling through vacuum. For transport through insulating or poorly conducting materials, charge transport can be enabled if the bandgap of the material is overcome by the energy provided by $V_{sd}$. Therefore, for charge transport through organic molecules, such as the linkers to the AuNPs or the surrounding biotemplate, one would expect nonlinear *I-V* characteristics in the range from −1 V to 1 V.[28,29] The temperature dependence for this charge transport is expected to be weak, because it is related to charge tunnelling along the molecules.

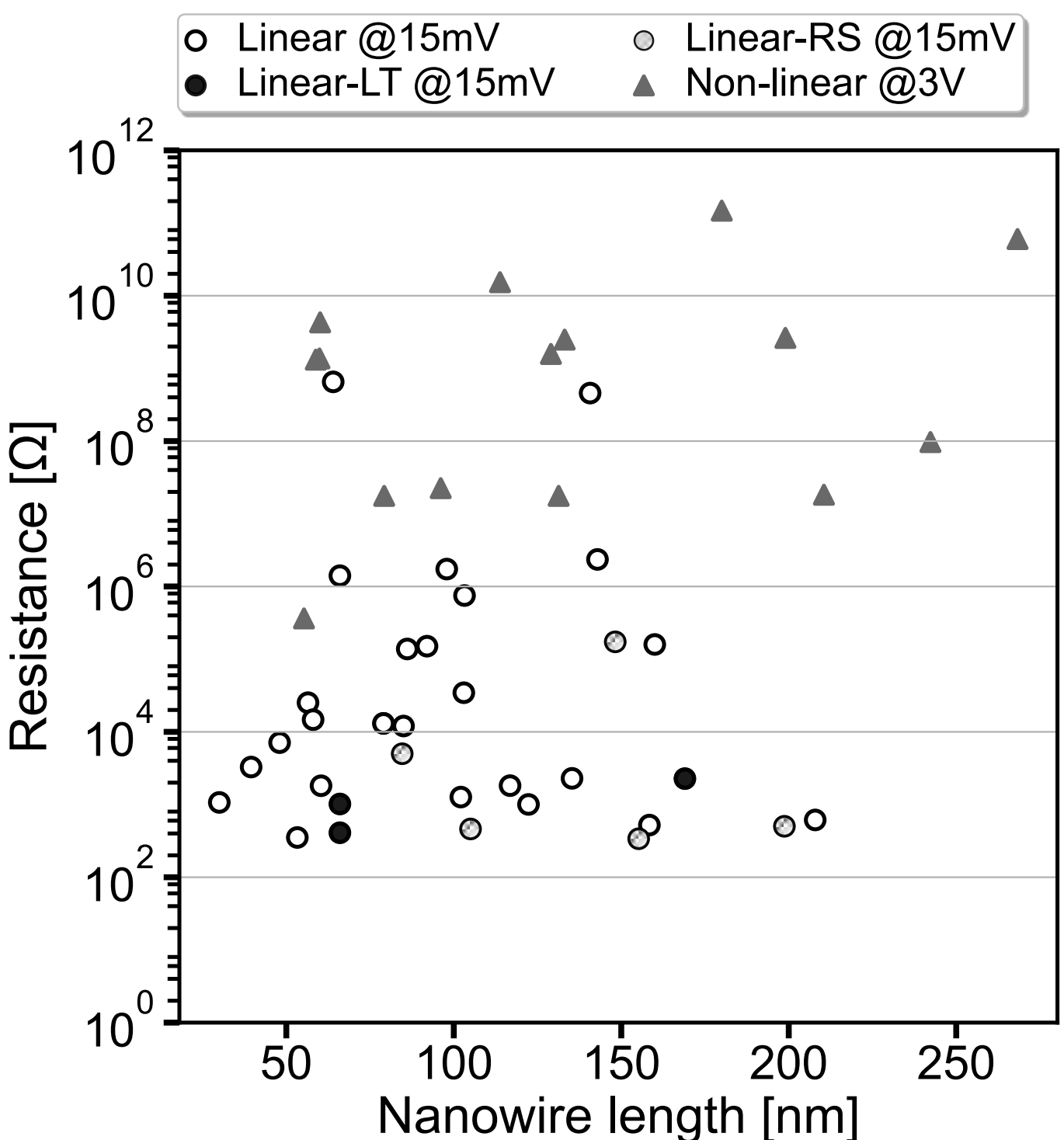

**Figure 3.** Resistance values as a function of length of the characterized AuNWs. AuNWs are classified according to the shape of their *I-V* curves and resistance values. NWs measured at low temperatures (Linear-LT) and NWs showing Resistive Switching (Linear-RS) are plotted with different symbols. Measurements were performed at room temperature and 1 × $10^{-5}$ mbar.

Unlike in other works reporting on microtubule-templated AuNWs, our work shows linear and nonlinear *I-V* characteristics in samples that have been processed using identical conditions within a single substrate. However, direct comparison is limited, because in previous experiments, contacting schemes and measurement conditions differed from those described in this work. Reported experiments either performed collective characterization by contacting multiple AuNWs,[27] or individual single AuNWs fabricated by a photoreduction reaction where gold ions were deposited on the microtubule outer surface.[26] After contacting a single AuNW (length of 2.5 µm, diameter between 80 nm to 100 nm) with platinum electrodes, a linear behavior was observed after repeated voltage application showing a resistivity of $7.3 \times 10^{-5}$ Ωm. Furthermore, the authors of this work comment on an improvement in the apparent coverage and continuity of the nanowire after fabricating the platinum electrodes and prior to the electrical measurements.[26] In contrast, our samples did not exhibit noticeable changes in AuNWs morphology after contact fabrication (**Figure 2**, first two columns). However, the electrical characterization exposed variations in the AuNWs resistance which cannot be directly attributed to contact-induced morphological changes. A clear example is the electric-shock failure, which predominantly occurs near the AuNWs midpoint, corresponding to the mechanically weakest region, rather than the contact interfaces (**Figure S2**). This behavior is likely caused by high voltages at wire sections with reduced width and Joule self-heating resulting from the high current density. Moreover, we observed a decrease in resistance for some AuNWs after the initial measurement, a behavior consistent with previous works.[23]

To further investigate this phenomenon, we analyzed the temperature-dependent charge transport of various linear AuNWs. Previous studies on similarly synthesized AuNWs, which were chemically grown from individual AuNPs using a biotemplating approach, have identified thermally activated hopping between nanoparticles as the primary transport mechanism.[18,24] Our measurements were carried out for both conductive ($R < 1$ kΩ) and resistive ($R > 10$ kΩ) AuNWs (**Figure 4** and **Figure S3**). After measuring at 300 K, without breaking the vacuum, the AuNWs were cooled down to 35 K and measured in incremental steps up to 300 K again. **Table 1** presents a showcase of three AuNWs that exhibit anomalous resistance changes, which cannot be explained solely by temperature effects. In all three cases, the resistance measured at 300 K after warm-up differs significantly from the initial value at 300 K. Specifically, we observed improved conductance in the first two AuNWs, with NW2 showing a resistance decrease of 63%. In contrast, NW3 displays a dramatic increase in resistance of nearly 900%. These results provide strong evidence that the observed behavior cannot be rationalized by simple thermal activation, as indicated by the unrealistically small activation energies extracted from Arrhenius fits (see fitting plots in the **Figure S4**).

**Table 1.** Electrical resistance of three AuNWs (NW1-NW3, length indicated in parentheses) measured at 300 K (initial), 75 K, and again at 300 K after warm-up, together with extracted activation energies $E_a$ restricted to a certain temperature range.

| | **NW1** (66 nm) | **NW2** (169 nm) | **NW3** (30 nm) |
|---|---|---|---|
| 300 K initial | 0.41 kΩ | 2.25 kΩ | 1 kΩ |
| 75 K | 0.85 kΩ | 0.81 kΩ | 11 kΩ |
| 300 K (warm-up) | 0.27 kΩ | 0.84 kΩ | 10 kΩ |
| $E_a$ | 10.8 meV | Metallic | 1.1 meV |

**Figure 4** presents the temperature-dependent *I-V* characteristics of the three representatives AuNWs listed in **Table 1** over the range of 35-300 K. A striking example is provided by NW2, where the observed whole temperature dependence cannot be rationalized within a conventional framework. Nevertheless, in the absence of transitions (75K – 300K) NW2 exhibits a stable metallic behavior (**Figure S4B**). This indicates that, between switching events, the AuNWs follow the expected conduction mechanism for such optimal metallic nanostructures.[18,23]

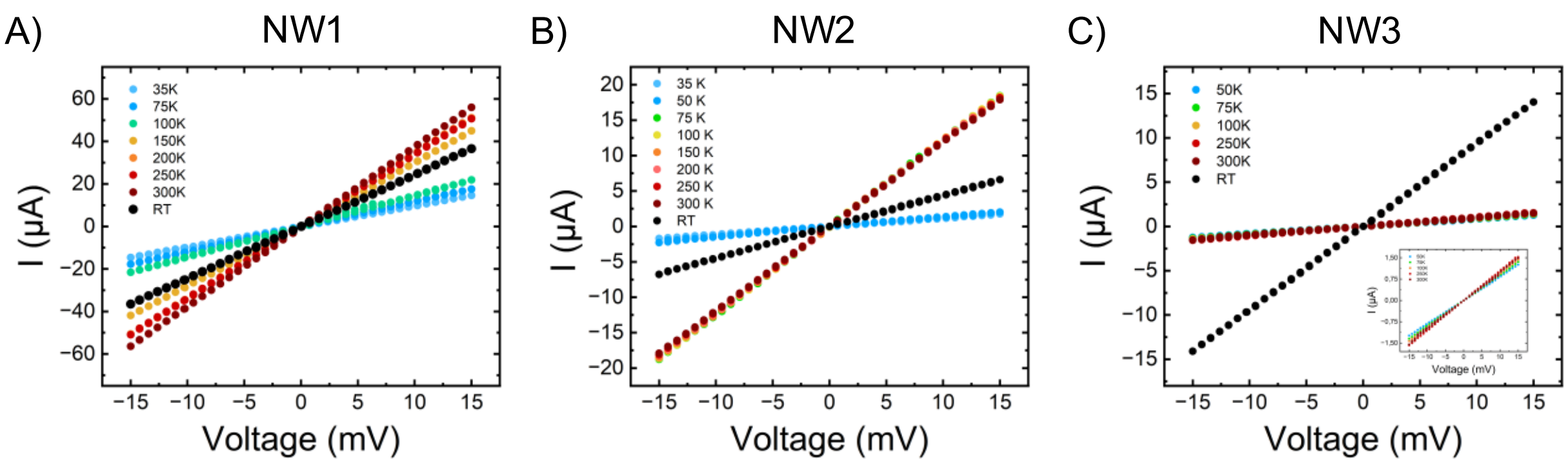

**Figure 4. A, B, C)** *I-V* sweeps of three individual microtubule-templated AuNWs (presented in Table 1) as a function of temperature in the range of 35-300 K. All AuNWs exhibit a nonlinear temperature dependence, revealing a deviation from purely metallic conduction.

To gain additional insight, we recorded such a transition for NW2 occurring at 75 K during an *I-V* sweep, where at a certain voltage, the current abruptly increases by approximately one order of magnitude (**Figure 5**). The initial resistance (R1) is 8.0 kΩ, whereas the final resistance state (R4) is 0.8 kΩ. Notably, during the first two voltage sweeps the current response is stable and reproducible. Without any external intervention, the AuNW undergoes this transition and subsequently remains locked in the new resistance state in the following measurements. Insets display the AuNW before electrical characterization (**Figure 5A**) and after the transition (**Figure 5C**). Post-characterization imaging reveals widening at the AuNW midpoint, suggesting local structural changes. These observations suggest the presence of tunable resistance states arising from structural reconfigurations under electrical stress, pointing toward the emergence of a new resistive switching (RS) mechanism intrinsic to pure metallic nanowires.

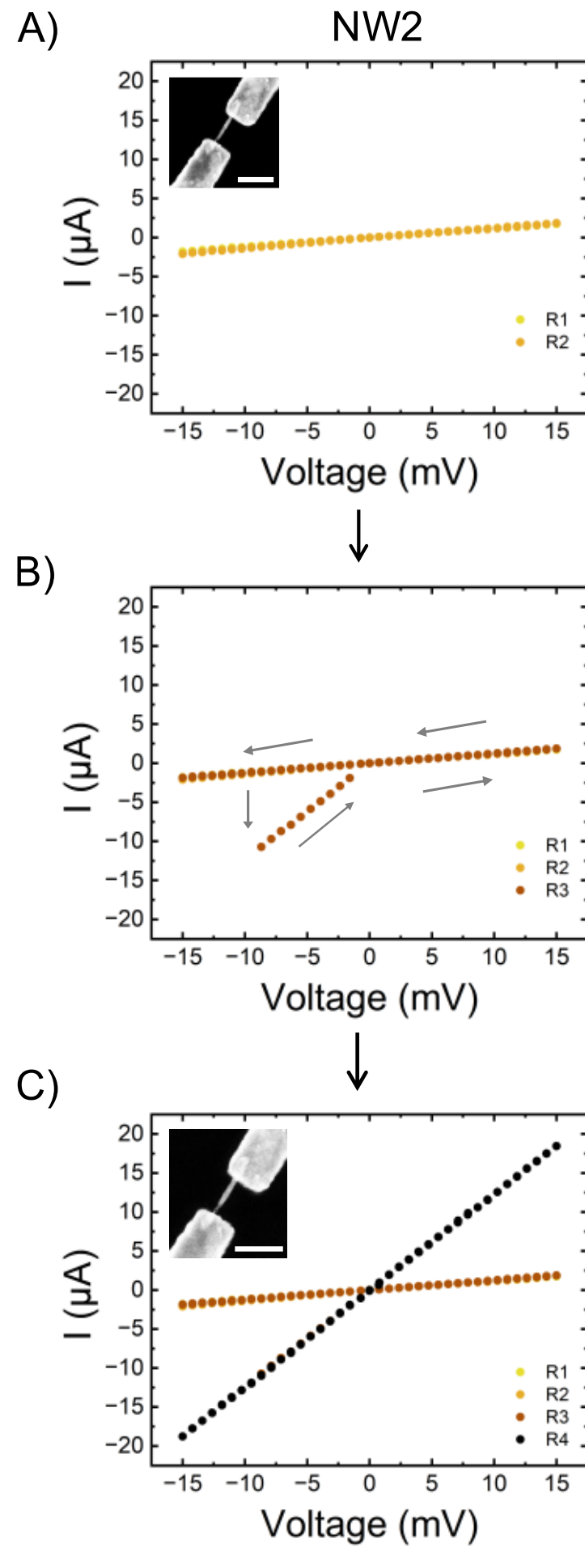

**Figure 5.** Sequentially repeated four *I-V* sweeps (R1-R4) of NW2 at 75 K. Panels **A, B, C)** present the consecutive measurements, illustrating an abrupt transition observed during R3. Insets in **A)** and **C)** show SEM images of the AuNW before and after electrical characterization, respectively. Scale bars = 200 nm.

Importantly, these switching events were observed across a broad range of resistance values and with similar frequency at both room temperature and low temperatures. The switching mechanism remains unchanged with temperature which is a strong indication that thermal effects, such as Joule heating, may contribute but are not the cause of the local rearrangements. Instead the data point toward electromigration as the governing mechanism, which occurs in nanoscale conductors at a high current density due to two main microscopic forces: (i) the high electric field, which acts on charged defects in the conductor at the nanoscale (direct force) and (ii) the momentum transfer from the electrons to defects (electron-wind force)[30]. The electric field ($E_{sd}$) in our AuNWs at a typical length of ~ 200 nm and $V_{sd}$ = 15 mV is $E_{sd} \approx 75$ kV m$^{-1}$. The current density flowing through the AuNW ( $j_{wire}$) depends on the diameter of the wire and the resistivity of the material. Since the resistance of the wires varies over orders of magnitude even for almost identical geometries (**Figure 3**), we can assume that $j_{wire}$ varies on a similar level. We therefore argue that the dominant force for the electromigration is the direct force caused by the large values of $E_{sd}$. This argument has important consequences on the operation of the AuNWs as reconfigurable switches. Unlike the wind-force, which always weakens the thinnest area of the wire because in this point $j_{wire}$ is largest, the direct force can lead to an increase as well as a decrease in resistivity due to the motion of defects in the external field.

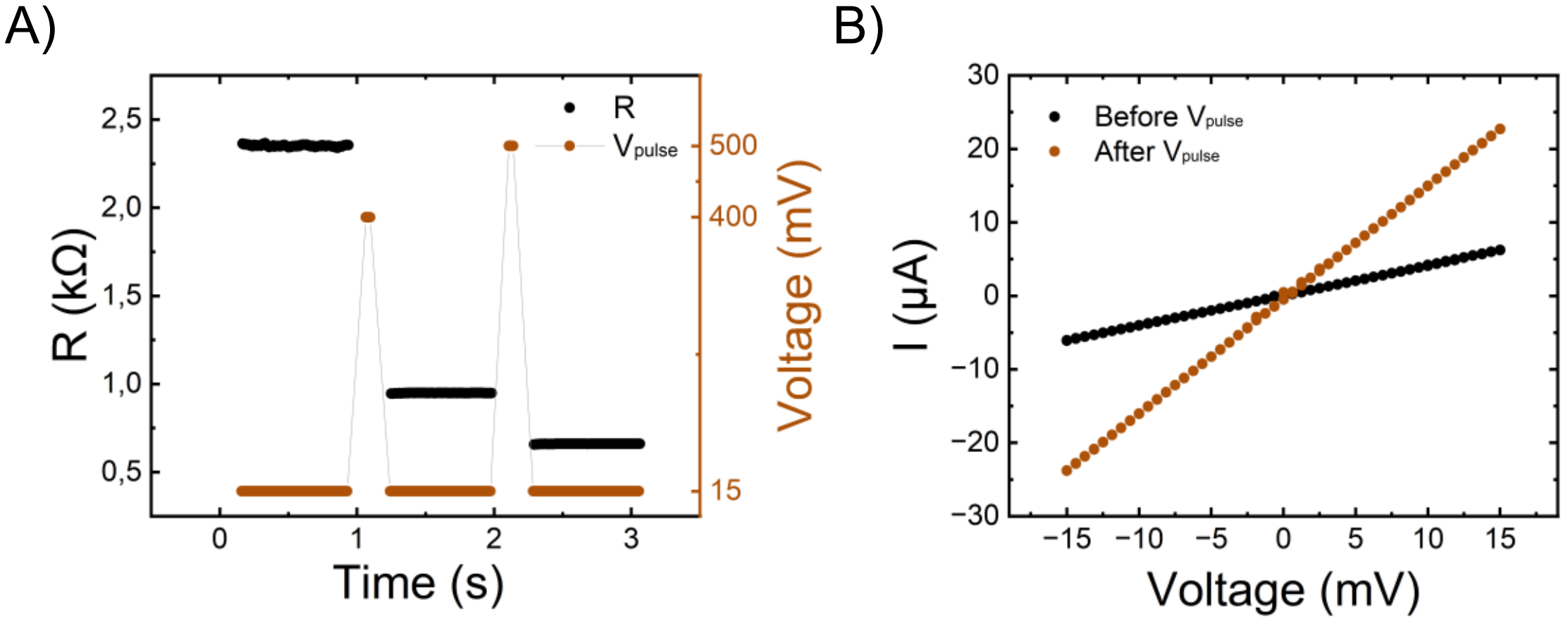

**Figure 6.** Pulse-bias-induced switching of a AuNW into new resistance states at 50 K. **A)** Resistance evolution upon application of two consecutive $V_{sd}$ pulses with varying pulse heights. **B)** *I-V* characteristics of the AuNW recorded before and after applying multiple single pulses, demonstrating a transition to a distinct conductive state.

Engineered millisecond pulses with peak voltages in the order of hundreds of mV induce changes in the AuNW resistances at 50 K, driven by localized structural reorganization (**Figure 6**). For voltage pulses with amplitudes below 400 mV, we did not observe resistance switching (**Figure S5**). The resistance of the AuNW changed from initially 2.3 kΩ to 0.95 kΩ after a pulse with $V_{sd}$ = 400 mV and to 0.66 kΩ after a pulse with $V_{sd}$ = 500 mV (**Figure 6A**). The resistance changes demonstrate that the electrical state of the AuNW can be dynamically tuned via controlled voltage pulses. Additional experiments revealed that the resistance can increase under similar pulsing conditions, indicating that atomic-scale rearrangements within the AuNWs are stochastic (**Figure S6**). Additionally, *I-V* measurements before and after a 400-mV pulse (**Figure 6B**) retain linear characteristics, confirming that metallic conduction remains the dominant charge transport mechanism while the AuNW undergoes resistance modulation. Together, these findings confirm the existence of tunable resistance states and reveal a previously unreported, intrinsic resistive switching mechanism in pure metallic nanowires.

## 3. Conclusion

We have characterized the electrical properties of AuNWs templated within the lumen of functionalized microtubules. The AuNWs exhibit metallic conduction, confirming the formation of continuous gold structures, yet display pronounced resistance variations under applied source-drain voltages. These changes are attributed to local structural reorganizations within the metallic lattice. As in other self-organized nanowires, both resistance increases and decreases were observed, reflecting local variations in the electric field along the wire. Crucially, our measurements demonstrate for the first time in such a system that resistance states can be actively tuned by voltage pulses. This novel reconfigurability positions microtubule-templated AuNWs as promising building blocks for neuromorphic electronics, where dynamic interconnectors are essential.

Their lateral geometry on wafer surfaces further facilitates integration into CMOS-compatible environments. Realizing their full potential will require reproducible control over resistance changes, achievable by (i) standardizing nanowire growth conditions and (ii) engineering contact geometries to tailor local field distributions. Together, these advances establish self-organized AuNWs as a versatile platform for reconfigurable nanoelectronic applications.

## 4. Experimental Section

**Microtubule polymerization and functionalization with the conjugates**

Guanosine-5'-($\alpha$, $\beta$)-methyleno triphosphate (GMP-CPP)-stabilized microtubules were polymerized and functionalized with AuNPs-Fab conjugates in a molar ratio of 3:1 (tubulin:conjugates), as described in [20]. After incubation, the microtubules were centrifugated for 15 min at 13.3 krpm to remove unbound gold nanoparticles and exchange the buffer to BRB80.

**AuNW synthesis and deposition on the $SiO_2$ substrates**

After pelleting and resuspending the functionalized microtubules in 40 µL of BRB80 buffer, 8 µL of this solution were added to a round vial containing 20.1 µL of BRB80 buffer to start the synthesis. Similarly to [20], the synthesis of the AuNWs was carried out with tetrachloroauric acid ($HAuCl_4$) (Merck) reduced by hydroxyl amine ($NH_2OH$) (Merck) in a 0.5:1 ratio. First, 2.9 µL of 1.54-mM $NH_2OH$ were added to the vial and gently mixed with a pipette for 1 min, then 9 µL of 1-mM $HAuCl_4$ were incorporated in the vial by mixing for another minute. Then, both reagents were added at the same time and the mixing continued for another minute. The last step was repeated to complete 3 elongation cycles, as described in [20]. All the AuNWs solution was deposited onto surface-activated $SiO_2$ substrates. After 30 min, the substrates were dipped into a 1:1 ethanol (VWR Chemicals) : dd$H_2O$ (Milli-Q, Merck) solution for 30 s. Then, the substrates were rinsed with 2-propanol (Merck) and dried with $N_2$.

**Nanofabrication and electrical characterization**

The layout for the electrical device characterization was designed and fabricated on $SiO_2$ substrates using electron beam lithography (Raith e-line Plus). Specifically, substrates were p-Si (100) with a 280-nm-thick oxide serving as the insulator layer. Before their use, these substrates were cleaned in an ultrasonic bath of acetone for 10 minutes plus 2 min in 2-propanol. A bilayer of EL11/PMMA-A4 electron beam resists was spin-coated and baked at 150 ºC for 10 min. The e-beam parameters for exposure included a 10-kV acceleration voltage, a 120-µm aperture size for contact pads, and a 30-µm diameter of the aperture for markers. The resists were subsequently developed in IPA/DI (7:3) and DI, each for 30 s. Creavac CREAMET 600 was used to deposit an adhesion layer of 10 nm Ti at a rate of 2 Å $s^{-1}$, followed by 100 nm Au at a rate of 5 Å $s^{-1}$ for the prefabricated layout; while for the individual contacts on top of the AuNWs, an adhesion layer of 5 nm Ti (2 Å $s^{-1}$) followed by a 50-nm Au layer (5 Å $s^{-1}$) were used. The final lithographic step consisted of an overnight lift-off process in acetone. Prior to microtubule-templated AuNW deposition via drop casting, the substrates were treated with an $O_2$ plasma at a flow rate of 7 sccm, input power of 200 W for 3 min (PICO, Diener Electronic-Plasma Surface Technology) to increase electrostatic negative charge and hydrophilicity of the surface. The random arrangement of NWs in the prefabricated layout enabled their individual identification and contacting. The AuNW coordinates were recorded via SEM imaging relative to the markers' arrays, loaded in the EBL software and used to perform another lithographic step to place tailored electrodes on top of the imaged AuNWs. Two identical $O_2$ plasma treatments at a flow rate of 5 sccm, input power of 200 W for 25 were performed to remove the organic template surrounding the AuNWs before electrode placement. SEM images were used to confirm the precise placement of the gold electrodes between the microtubule-templated AuNWs. This lithographic approach was adapted according to the wire lengths and resulted in a wide range of electrode-electrode distances. Two terminal *I-V* measurements on 50 individual AuNWs for the electrical characterization were carried out in darkness and vacuum ($1 \times 10^{-5}$ mbar base pressure) using a Keithley 2400. Tungsten tips of 25-µm diameter were placed on the gold contact pads. For temperature-dependent electrical characteristics, a liquid helium continuous flow cryostat system cooled the samples in a range between 35 K to 300 K ($1 \times 10^{-7}$ mbar base pressure). Fixed voltage measurements were carried out by applying a constant 15 mV bias and recording the current as a function of time. Pulse-induced resistance modulation experiments were performed by applying consecutive voltage pulses with amplitudes from 100 mV to 500 mV and pulse widths of 40 ms, followed by *I-V* sweeps to monitor the state evolution. *I-V* measurements were performed in a continuous voltage sweep from 0 mV to 15 mV, then 15 mV to −15 mV, and back to 0 mV. For linear wires, resistance was calculated through a linear fit for the obtained current. For wires with nonlinear behavior, resistance was determined by the current at the maximum applied voltage.

## Acknowledgments

B.R.B., B.P.R., S.D. and A.E. acknowledge financial support funded by the Deutsche Forschungsgemeinschaft (DFG, German Research Foundation) - GRK 2767 - Project number 451785257. B.R.B, and A.E. acknowledge support by the Nanofabrication Facilities Rossendorf (NanoFaRo) at the IBC. The authors thank A. Worbs for TEM specimen preparation. Furthermore, the use of the HZDR Ion Beam Center TEM facilities and the funding of TEM Talos by the German Federal Ministry of Education and Research (BMBF, grant No. 03SF0451) in the framework of HEMCP are acknowledged.

## Conflict of Interest and Data Availability Statements

The authors declare that the data supporting the findings of this study are available within the paper and its Supplementary Information files. If raw data files are needed in another format, they are available from the corresponding author upon reasonable request.

The authors declare no conflicts of interest regarding this manuscript.

# Supporting Information

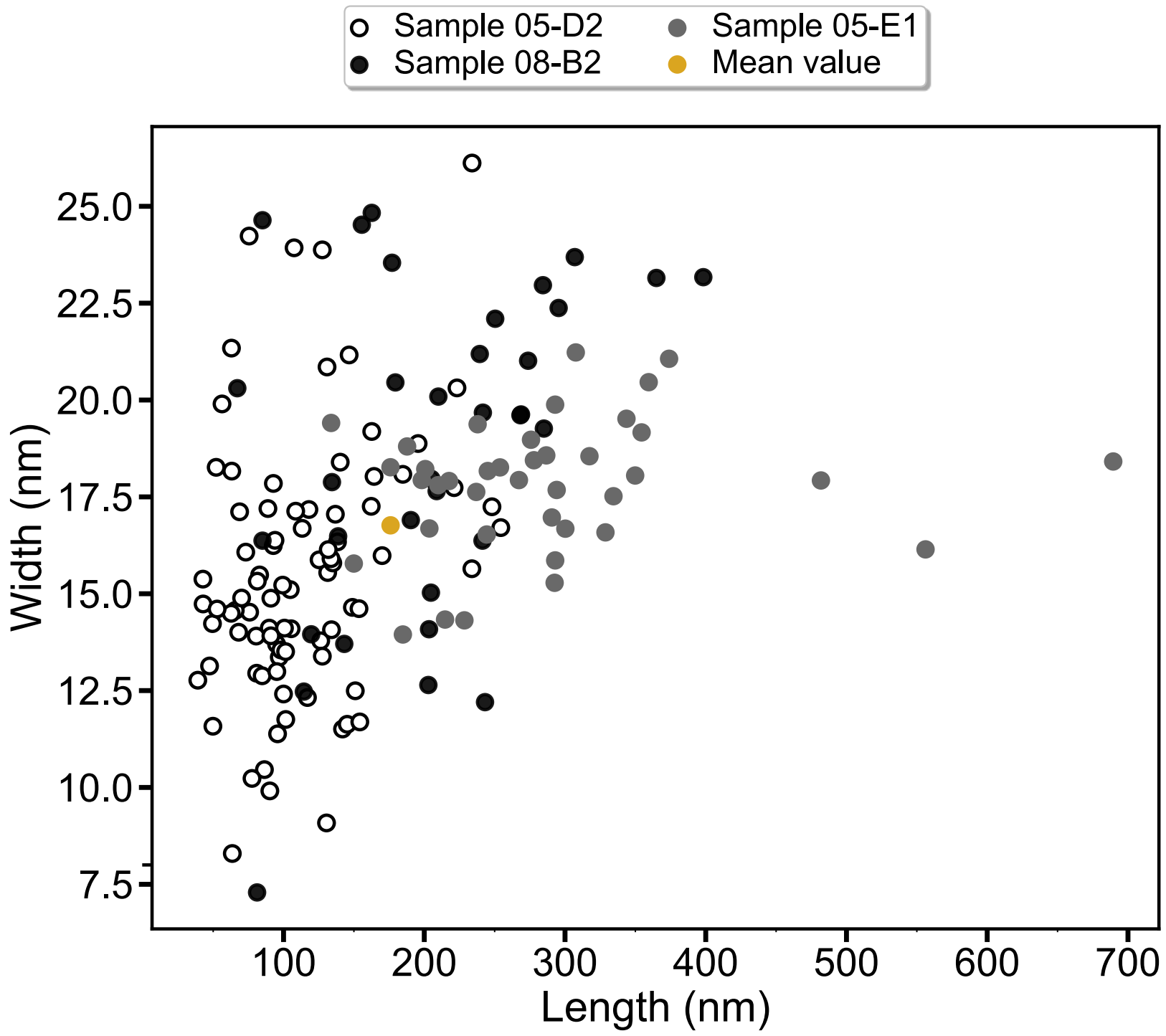

**Figure S1.** Scatter plot showing the length and width of the AuNWs deposited on silicon wafers. The average length and width are 176 nm and 16.7 nm, respectively.

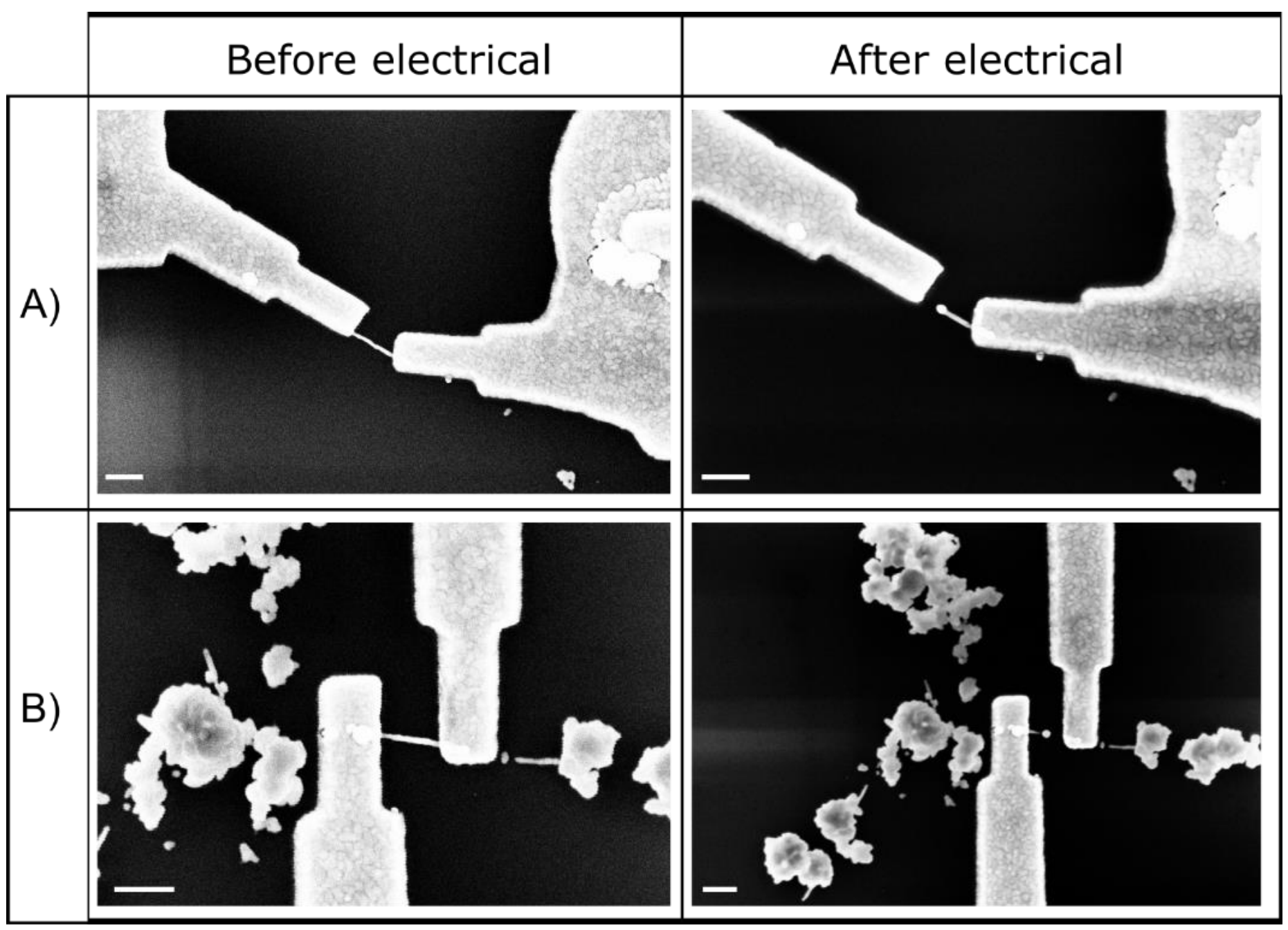

**Figure S2. A, B)** Two NWs before (left) and after (right) electrical measurements. The wires were molten during the measurements, and the ends of the broken wires form metallic spheres in the central part. Scale bars = 200 nm.

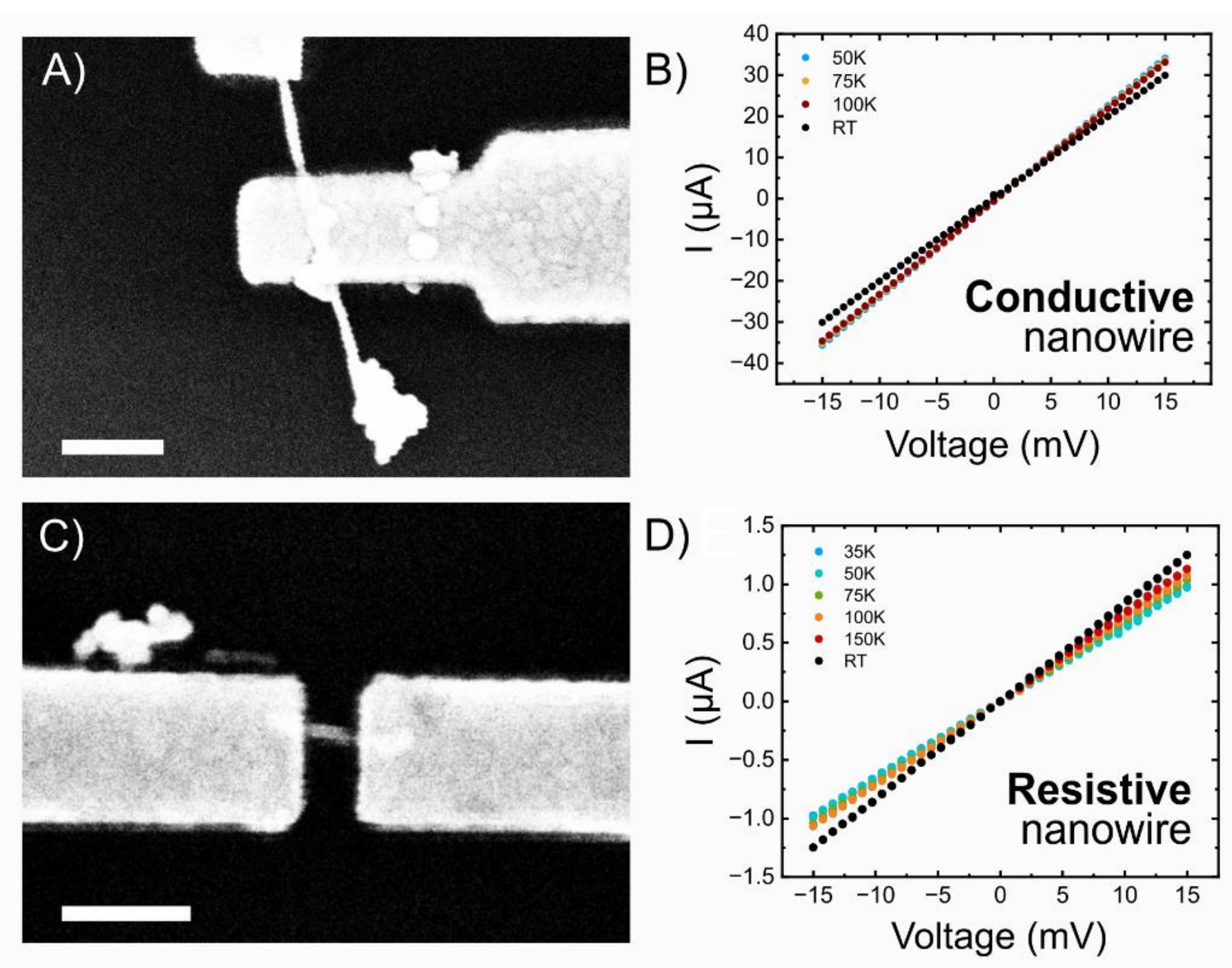

**Figure S3:** Examples of AuNWs in different resistance regimes over temperature. SEM images from **A)** a conductive and **C)** a resistive AuNW and their corresponding *I-V* curves, **B)** and **D)** respectively, at different temperatures in the range of 35-300 K. The conductive nanowire exhibits temperature-dependent metallic behavior, and the resistive wire exhibits thermally activated behavior. Scale bars = 200 nm.

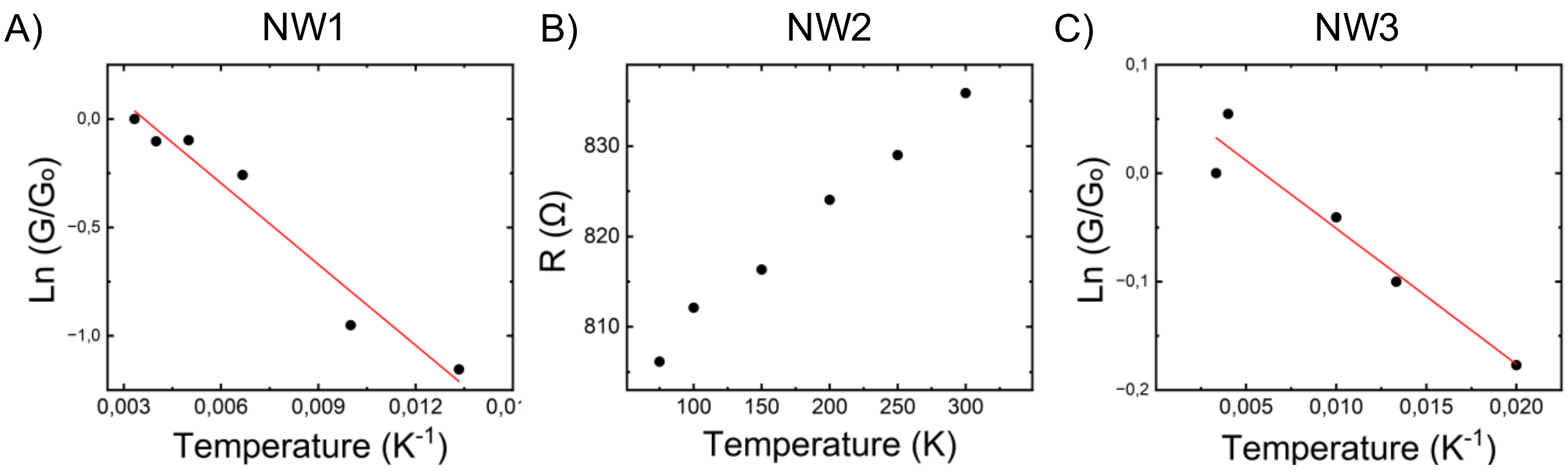

**Figure S4. A)** Normalized logarithm of Arrhenius plot for NW1 restricted to the temperature regime of 75-300 K, with a linear fit shown as a red solid line. **B)** Resistance versus temperature plot for NW2 in the range of 75-300 K displaying a metallic behavior. **C)** Normalized logarithm of Arrhenius plot for NW3 restricted to the temperature regime of 50-300 K, with a linear fit shown as a red solid line.

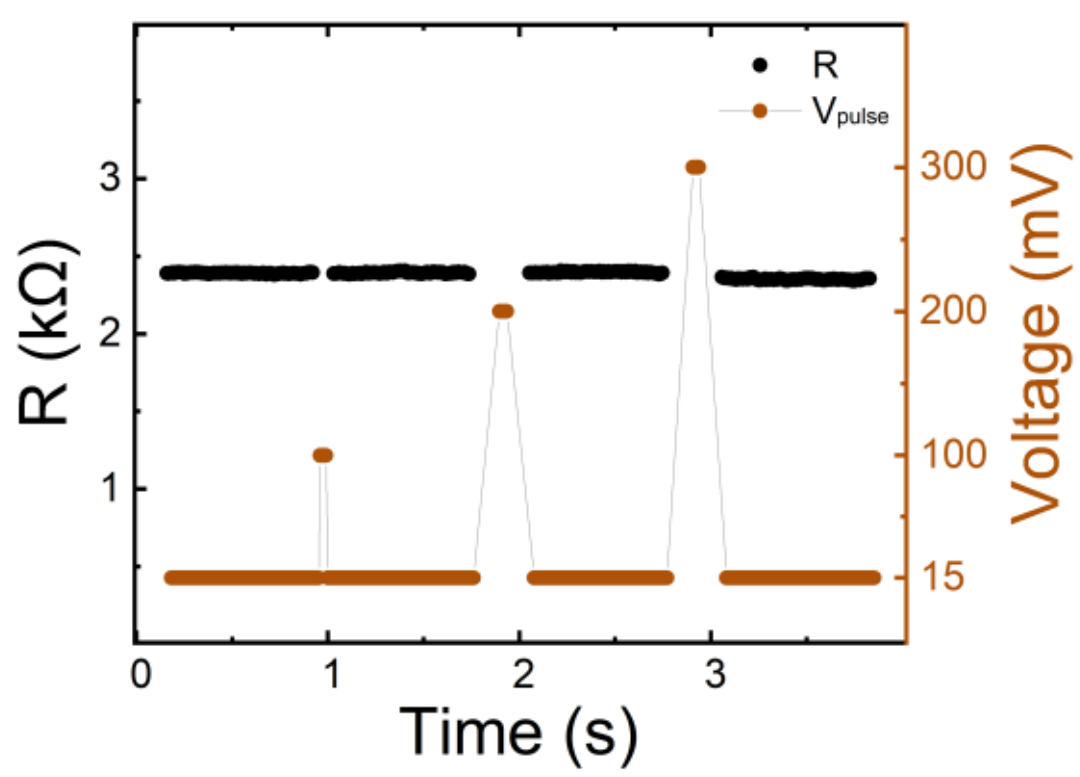

**Figure S5.** Low voltage pulses below 400 mV (range 100 to 300 mV) at 50 K did not lead to changes in the resistance.

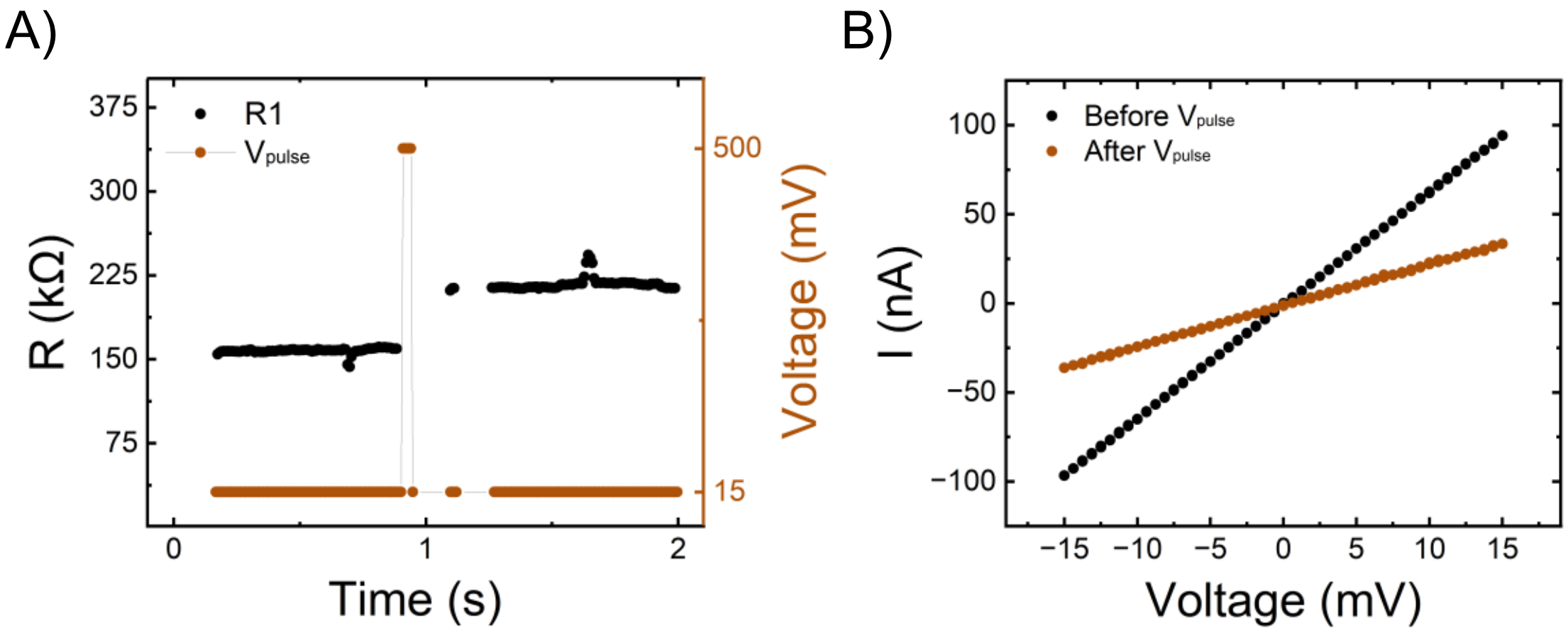

**Figure S6.** Pulse-induced transitions to a higher resistance state at RT. **A)** Resistance modification a 500 mV $V_{sd}$ pulse and **B)** *I-V* curves of the AuNW before and after applying multiple single pulses showing a switch in the resistance.

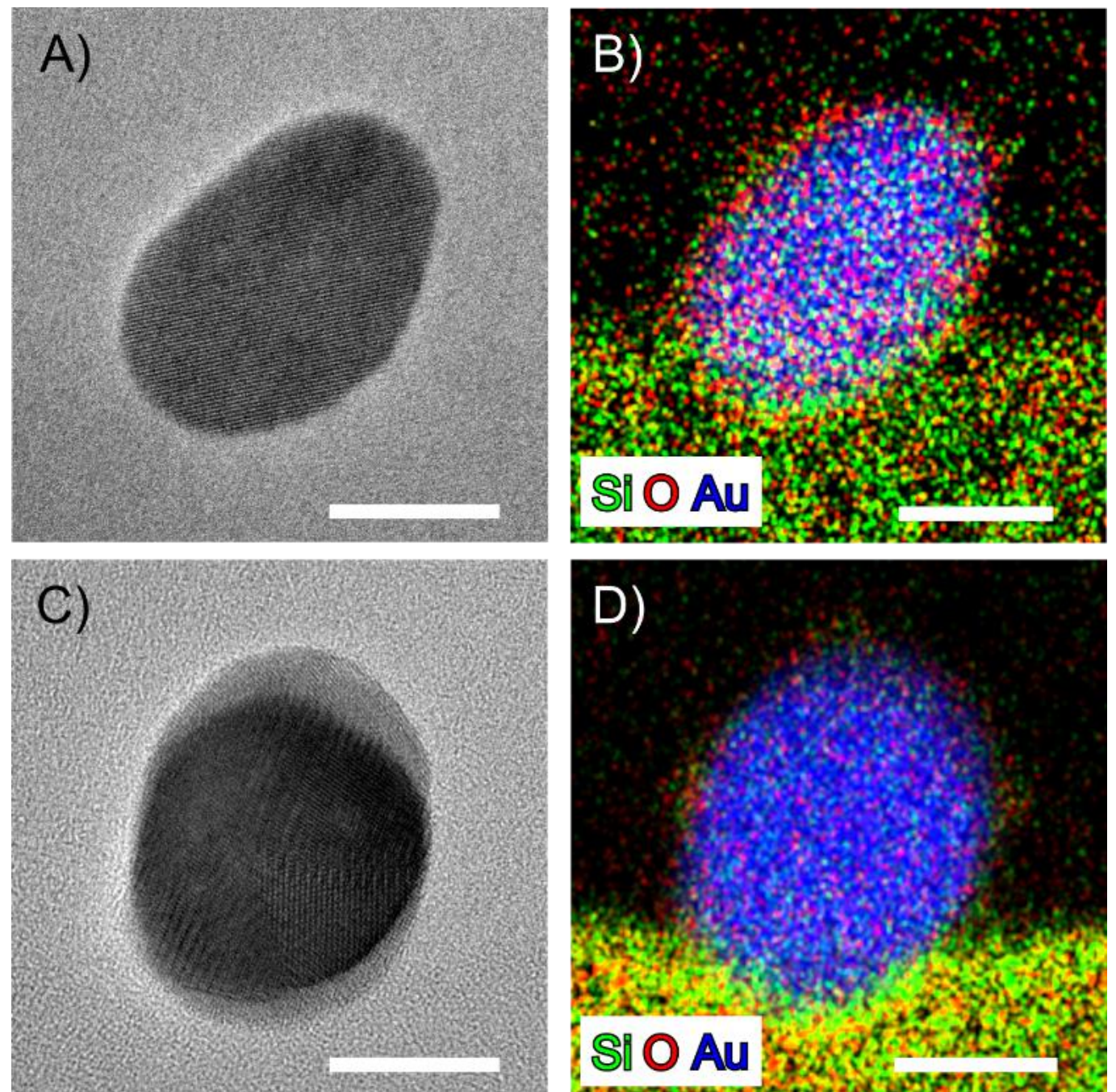

**Figure S7.** Enlarged views of the cross-sectional TEM images **A)** and **C)** of two different AuNWs and their corresponding element composition analyses, **B)** and **D)** respectively. Scale bars = 10 nm.

## Table of Contents

# Intrinsic Resistive Switching in Microtubule-Templated Gold Nanowires for Reconfigurable Nanoelectronics

The scaling limits of CMOS transistors require alternative, dynamically reconfigurable nanoscale devices. This study demonstrates the intrinsic resistive switching in pure gold nanowires templated by microtubules. Continuous nanowires with local inhomogeneities enable electromigration-driven, multistate resistance modulation under voltage pulses, while retaining metallic conductivity. Their high aspect ratio, and CMOS-compatible processing position them as an ideal platform for reconfigurable nanoelectronic interconnects.

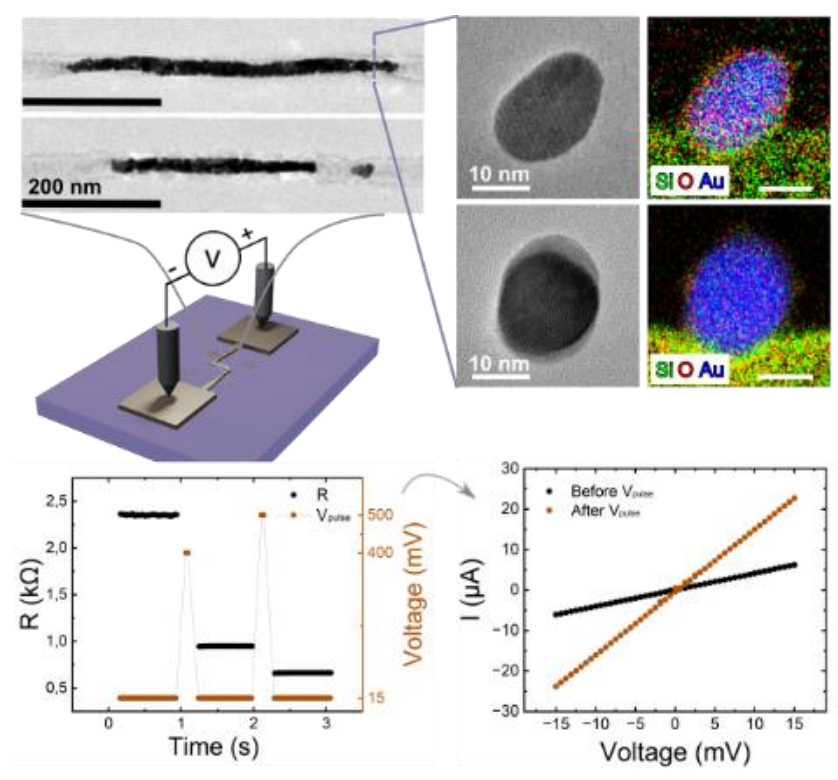